\documentstyle[twoside,fleqn,espcrc2,psfig]{article}

\title{Scaling Structures in Four-dimensional Simplicial Gravity
\thanks{presented by H.S.Egawa}}

\author
{
H.S.Egawa \address{
Department of Physics, Tokai University 
Hiratsuka, Kanagawa 259-12, Japan}, 
T.Hotta $^{\,\, {\rm b}}$, 
T.Izubuchi \address{
Department of Physics, University of Tokyo 
Bunkyo-ku, Tokyo 113, Japan}
N.Tsuda \address{
National Laboratory for High Energy Physics (KEK), 
Tsukuba 305, Japan}
and 
T.Yukawa $^{\,\, {\rm c,}}$  
\address{
Coordination Center for Research and Education, 
The Graduate University for Advanced Studies, Hayama-cho, 
Miura-gun, Kanagawa 240-01, Japan}
}

\begin{document}
\begin{abstract}
Four-dimensional(4D) spacetime structures are investigated using 
the concept of the geodesic distance in the simplicial quantum 
gravity. 
On the analogy of the loop length distribution in 2D case, 
the scaling relations of the boundary volume distribution in 4D 
are discussed in various coupling regions $i.e.$ strong-coupling phase, 
critical point and weak-coupling phase.
In each phase the different scaling relations are found.
\end{abstract}                                                         
\maketitle

%
\section{Introduction}
Simplicial gravity has witnessed a remarkable development toward 
quantizing the Einstein gravity.
This development started with 2D simplicial gravity 
and has now reached the point of subjecting to simulate 4D 
case\cite{Agis_Migd,Scal_4DQG,Phase_4D,4DDT} 
about the analysis of fractal dimensions, 
minbu, scaling relations for the loop length distribution and 
the curvature distribution. 
The aim of this paper is to investigate 4D Euclidean 
spacetime structures using the concept of the geodesic 
distance. 
It is very important that the scaling relations have been obtained 
in 2D case. 
Therefore, on the analogy of the loop length distribution(LLD) in 
2D\cite{KKMW}, the scaling relations in 4D are discussed.
Actually we measured the boundary volume distribution(BVD) for various 
geodesic distances in 4D dynamically triangulated(DT) manifold, 
in analogy to LLD. 
In order to discuss the scaling relations, we assume that the scaling 
variable $x$ has a form $x=V/D^{\alpha}$, where $V$, $D$ and $\alpha$ 
denote the each boundary(cross section) volume, the geodesic distance 
and scaling parameter, respectively.
Hagura ${\it et \; al.}$ argue the scaling properties of the surface area 
distributions in 3D case by the same analysis as we employ in 4D case(in 
these proceedings).
%
\section{The model}
We use the lattice action of 4D model with the $S^{4}$ topology 
corresponding to the action as  
$S = - \kappa_2 N_2 + \kappa_4 N_4$,  
where $N_i$ denotes the total number of $i$-simplexes.
The coupling $\kappa_2$ is proportional to the inverse bare Newton 
constant and the coupling $\kappa_4$ corresponds to a lattice 
cosmological constant. 
For the dynamical triangulation model of 4D quantum gravity, 
we consider a partition function of the form, 
$Z(\kappa_2, \kappa_4) = \sum_{T(S^4)} e^{-S(\kappa_2, \kappa_4, T)}$.
We sum up over all simplicial triangulations $T(S^4)$. 
In practice, we have to add a small correction term, 
$\Delta S = \delta \kappa_4 (N_4 - N_4 ^{(target)})^2$, to the 
action in order to suppress the volume fluctuations from the target 
value of $4$-simplexes $N_4 ^{(target)}$ and we have used 
$\delta = 0.0005$.
%
\section{Numerical Simulations and Results}
\begin{figure}
\centerline{\psfig{file=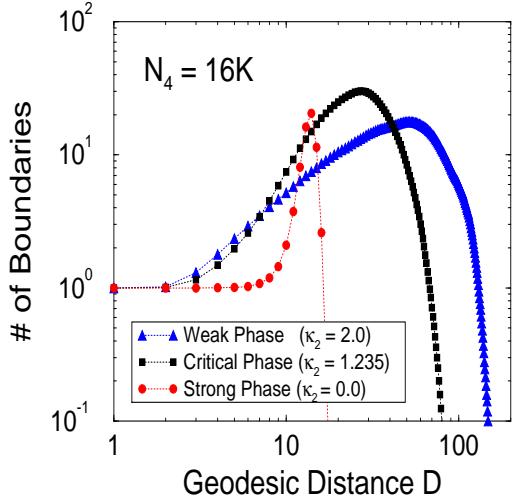,height=7cm,width=7cm}} 
\vspace{-0.5cm}
\caption
{
Number of Boundaries plotted versus geodesic distances $D$, with the 
double-log scales.
}
\label{fig:Number_Boundary_16K}
\end{figure}

We define $N_{b}(D)$ as the number of boundaries at the geodesic 
distance $D$ from a reference $4$-simplex in the $4$D DT manifold averaged 
over all $4$-simplexes.
Fig.\ref{fig:Number_Boundary_16K} shows the distributions of $N_{b}(D)$ 
for the typical three coupling strength with $N_{4} = 16K$.
In the strong coupling limit($\kappa_{2}=0$), the only one boundary that is 
identified as the mother universe exists almost all the distances(see Fig.1), 
which means that the mother universe is a dominant structure.
The branching structures are highly suppressed, which shows characteristic 
properties of the ``crumpled manifold'', which is similar to the case 
observed in $2D$ manifold.
On the other hand, in the weak coupling phase(for example, we chose 
$\kappa_{2}=2.0$), we observe the growth of the branches until $D \sim 60$ 
and can reasonably extract the relation $N_{b}(D) \propto D$ in the region 
$3 \leq D \leq 30$.
Then we call this manifold as the ``elongated manifold''.
%
\begin{figure}
\centerline{\psfig{file=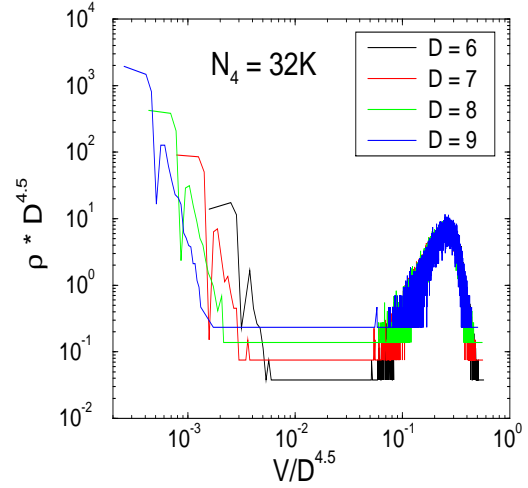,height=7cm,width=7cm}} 
\vspace{-0.5cm}
\caption
{
BVD plotted versus a scaling variable $x = V/D^{4.5}$, with the double-log 
scales in the strong coupling limit: $\kappa_{2} = 0$ and $N_{4}=32K$.
}
\label{fig:BVD_Strong}
\end{figure}
\begin{figure}
\centerline{\psfig{file=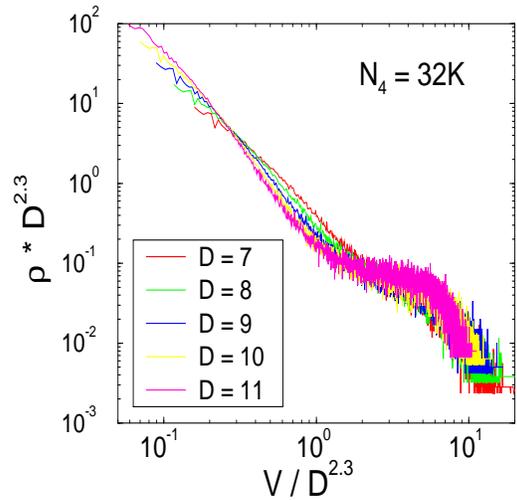,height=7cm,width=7cm}} 
\vspace{-0.5cm}
\caption
{
BVD plotted versus a scaling variable $x = V/D^{2.3}$, with the double-log 
scales at the critical point: $\kappa_{2}=1.2581$ and $N_{4}=32K$.
}
\label{fig:BVD_Critical}
\end{figure}
\begin{figure}
\centerline{\psfig{file=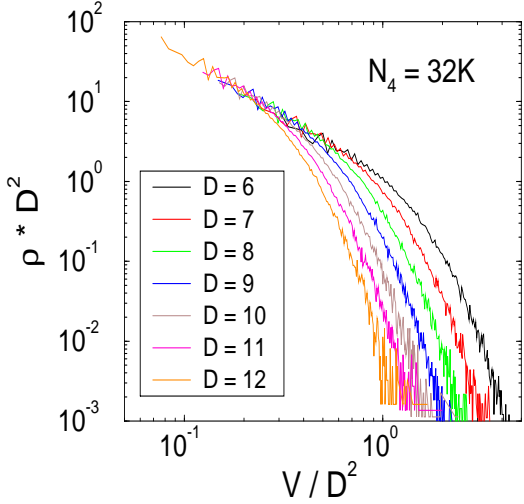,height=7cm,width=7cm}} 
\vspace{-0.5cm}
\caption
{
BVD plotted versus a scaling variable $x = V/D^{2}$, with the double-log 
scales in the weak coupling phase: $\kappa_{2} = 2.0$ and $N_{4}=32K$.
}
\label{fig:BVD_Weak}
\end{figure}
\subsection{The strong coupling phase}
Fig.\ref{fig:BVD_Strong} shows BVD, $\rho(x)$, with $x = V/D^{4.5}$ as a 
scaling variable in the strong coupling region, $\kappa_{2} = 0$, with 
$N_{4}=32K$ while the fractal dimension($d_{f}$) reaches about $5.5$ which 
yet increases with the volume in our simulation size.
In terms of this variable the mother universe shows scaling relation and 
distributes like a Gaussian distribution.
We can be fairly certain that in the strong coupling phase the scaling 
parameter $\alpha$ of mother universe satisfies the relation 
$d_{f} = \alpha + 1$, and the manifold resembles a $d_{f}$-sphere
($S^{d_{f}}$). 
There seems to be a scaling property with respect to BVD 
with $x = V/D^{d_{f}-1}$ as a scaling variable.
%
\subsection{The critical point}
Next, the data near the critical point are shown in 
Fig.\ref{fig:BVD_Critical} for various geodesic distances with $N_{4} = 32K$.
At this point the fractal dimension reaches about $3.5$ and yet increasing 
for larger values of $N_{4}$.
We must draw attention to the double peak structure on the critical point
\cite{BBKP_4DQG}.
Therefore, we measure the boundary volume distribution on both peaks, 
and obtain the more clear signal of the distribution of the mother universe 
on the peak which is close to the strong coupling phase.
We observe the scaling properties for the mother universe $\rho(x)$ with 
$x=V/D^{2.3}$ in Fig.\ref{fig:BVD_Critical}.
In order to discuss the universality of the scaling relations, we assume the 
distribution function in terms of a scaling parameter $x = V/D^{2.3}$ as 
$\rho(x) = a_{0} \frac{1}{D^{2.3}} x^{a_{1}}e^{-a_{2}x}$,
where $a_{0}$, $a_{1}$ and $a_{2}$ are some constants.
Then we can calculate the fractal dimension from $\rho(x,D)$, 
$
\lim_{x \to \infty} \int_{v_{0}}^{\infty} dV \; V \; \rho(x,D) = V^{(4)}(D) 
\sim D^{d_{f}},
$
where $v_{0}$ denotes the cut-off volume and $V^{(4)}$ denotes the total 
volume of the 4D manifold.
If $a_{1} > -2$ this integration is convergent and gives a finite fractal 
dimension.
We can extract the function of $\rho(x,D)$ from Fig.\ref{fig:BVD_Critical}, 
and find $a_{1} \simeq 0.5$ for the distribution of mother volume and 
$a_{2} \simeq 3.0$.
Furthermore we investigate the $N_{4}=64K$ case and obtain the same scaling 
behavior as that of $N_{4}=32K$ case except $d_{f} \simeq 4.0$ and scaling 
parameter $\alpha \simeq 3.0$.
These results lead to the conclusion that the distributions of the baby 
universes show no scaling behavior.
On the other hand, the distribution of the mother universe shows the scaling 
relation and is universal($i.e.$ it does not depend on the lattice 
cut-off($v_{0}$)). 
%

\subsection{The weak coupling phase}
Finally, we show the data in the weak coupling phase($\kappa_{2}=2.0$) 
within which the fractal dimension reaches about $2.0$.
Fig.\ref{fig:BVD_Weak} shows BVD with $x = V/D^{2}$ as a scaling variable. 
We can safely state 
$\rho (x) \times D^{2} \propto x^{-2.0} e^{-x}$.
In this phase, the dynamically triangulated manifold consists of widely 
expanding like branched polymers and we cannot observe the mother universe 
at all.

\section{Summary and discussions}
On the analogy of LLD in 2D case, the scaling relations in 4D are 
discussed for the three phases.
BVD, $\rho(x,D)$, at geodesic distance $D$ gives us some basic scaling 
relations on the ensemble of Euclidean space-times described by the 
partition function $Z(\kappa_2, \kappa_4)$.

In the strong coupling limit $\kappa_{2}=0$ we find that the mother part 
of BVD, $\rho(x,D)$, scales trivially with 
$x=V/D^{d_{f}-1}$ as a scaling variable. 
There is fairly general agreement that the 4D DT manifold seems to be a 
$d_{f}$-sphere($S^{d_{f}}$).
What is important is that this scaling property for the mother universe 
changes gradually into the scaling relation of that of the critical point. 
The fluctuations of the spacetime growth with 
$\kappa_{2} \to \kappa_{2}^{c}$.
In 2D, the baby loop and the mother loop show scalings with the same 
parameter($x=L/D^{2}$).
However, LLD of the baby loops is depend on the lattice cut-off and we 
think that it is not universal. 
At the critical point in 4D case we have obtained the similar BVD. 
However, we have a different scaling parameters
($x=V/D^{2.3}$ with $N_{4}=32K$ and $x=V/D^{3.0}$ with $N_{4}=64K$) 
from 2D case for the mother universe.
Furthermore, BVD of the baby universes seems to be non-universal. 

In the weak coupling phase(see Fig.\ref{fig:BVD_Weak}) we have obtained the 
elongated manifolds, in other words, branched polymers.
In this phase no mother universe exists and BVD of the baby universes 
shows that the scaling relation is not universal.

The results of this paper is the first step to research the universal 
scaling relations in 2, 3 and 4D on simplicial 
quantum gravity. 
The results of simulations can be regarded as the possibility that we may 
have 4D quantum gravity as the generalized DDK model. 
\vspace{-0.3cm}
\begin{center}
-Acknowledgment-
\end{center}
\vspace{-0.3cm}
We would like to thank H.Kawai, N.Ishibashi, S.R.Das, J.Nishimura 
and H.Hagura for fruitful discussions. 
Some of the authors (T.H., T.I. and  N.T.) were supported by a Research 
Fellowships of the Japan Society for the Promotion of Science for Young 
Scientists.


\end{document}